\documentstyle[prb,aps,twocolumn]{revtex}
\input psfig
\begin{document}

\author{ N.~I.~Lundin$^1$, L.~Y.~Gorelik$^{1,3}$, R.~I.~Shekhter$^1$, V.~S.~Shumeiko$^{2,3}$ and M.~Jonson$^1$ }
\title{\bf
\vspace*{-10mm}
Microwave controlled phase coherent transport in mesoscopic S-N-S structures}
\address{$^{(1)}$Department of Applied Physics,  \\
Chalmers University of Technology and G{\"o}teborg University, 
SE-412 96 G{\"o}teborg, Sweden\\
$^{(2)}$Department of Microelectronics and Nanoscience,\\
 Chalmers University of Technology and G{\"o}teborg University, 
SE-412 96 G{\"o}teborg, Sweden\\
 $^{(3)}$B. Verkin Institute for Low Temperature Physics and Engineering,
\\ 
National Academy of Sciences of Ukraine, 310164 Kharkov, Ukraine\\{(\rm \today)}\\{\parbox{14cm} 
{\small 
We show that transport through a superconducting quantum point contact biased at subgap voltages is strongly affected by a microwave field. The subgap current is increased by several orders of magnitude. Quantum interference among resonant scattering events involving photon absorption is reflected as an oscillating structure in the I-V curve. We also discuss how the same interference effect can be applied  for detecting weak electromagnetic signals up to the gap frequency, and how it is affected by dephasing and relaxation. }}}
\maketitle
\newcommand{\p}{{\bf u}}
\newcommand{\pccc}{{\bf u}}
\newcommand{\pcc}{{\bf u}}
\newcommand{\be}{\begin{eqnarray}}
\newcommand{\ee}{\end{eqnarray}}
\newcommand{\bup}{b^{+}}
\newcommand{\ffiup}{\p^{+}}
\newcommand{\ffiner}{\p^{-}}
\newcommand{\ffi}{\vec{\varphi}}
\newcommand{\bner}{b^{-}}
\newcommand{\sx}{\sigma_x}
\newcommand{\sy}{\sigma_y}
\newcommand{\sz}{\sigma_z}
\newcommand{\sxxx}{\sigma_1}
\newcommand{\syyy}{\sigma_2}
\newcommand{\szzz}{\sigma_3}
\newcommand{\sxx}{\tau_x}
\newcommand{\syy}{\tau_y}
\newcommand{\szz}{\tau_z}
\section{Introduction}

The electrical properties of small sized conductors at low temperatures 
are significantly affected by the finite spacing of electron energy levels,
resulting in a number of mesoscopic phenomena. A few examples of equilibrium 
or near equilibrium mesoscopic effects are magnetic moment
fluctuations in quantum dots and rings as well as 
universal conductance fluctuations
in quantum point contacts and nanowires.\cite{Altshuler,Datta} 
The opposite limit, of strong non-equilibrium 
behavior, appears for example when external fields produce 
significant redistribution
of electrons between quantized levels or modify the quantum states.
In this limit coherent time dependent dynamics
in mesoscopic conductors can be studied. Two possible approaches are 
nonlinear transport phenomena in strong electrical fields and optical
interlevel transitions under infrared irradiation.

An interesting possibility for coherent time dependent dynamics 
appears in mesoscopic structures where
the position of quantized electronic levels slowly (adiabatically)
change in space or time. Under this condition
external alternating fields will produce
a strong  resonant inter-mode coupling which will take place only
in the vicinity of certain points in space  or at certain times when
the resonance condition (the inter-level distance coincides
with the energy, $\hbar\omega$, of the external field) is satisfied~\cite{HN,somer}.
The adiabatic dynamics of the energy levels together with the well defined
inter-mode coupling events results in 
a novel type of non-equilibrium mesoscopic phenomena.
In Ref.~\onlinecite{prla} a ballistic channel with varying width is
studied. It is shown that interlevel transitions induced by infrared radiation
may have a similar effect as impurity scattering and significantly 
influence transport through the channel.
In this system, impurity scattering phenomenon like 
quantum localization of conducting electrons, and
phenomenon foreign to impure systems like 
resonant backscattering of electrons~\cite {prlf} will be present.
As an example of a non-equilibrium
mesoscopic phenomenon induced by inter-mode coupling
events localized in time one can consider how energy is stored in 
an disordered mesoscopic ring subject to a time-dependent magnetic flux.
The interference pattern which appears in this system as a result of 
the different temporally localized inter-mode transitions determines 
the various pictures of coherent energy accumulation that 
has been discussed in Ref.~\onlinecite{GT,prlg}.
Another very interesting subject to study is coherent dynamics 
of temporally localized inter-mode transitions in ballistic
Josephson junctions subject to infrared irradiation. This is the
subject to be addressed here.
\begin{figure}
\centerline{\psfig{figure=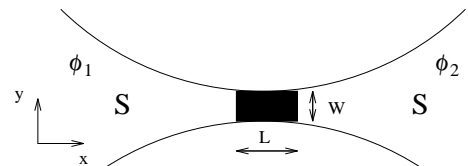,width=6cm}}
 \caption{ \label{fig:junction}
Single mode adiabatic superconducting microconstriction with normal region of length L, width W and transmission coefficient D. The phase difference over the
junction is defined as $\phi=\phi_2-\phi_1$. } 
\end{figure}
It is well known that the current through a voltage-biased
superconducting quantum point contact (SQPC) is carried by localized states. 
These states, called Andreev states, 
are confined to the normal region of the
contact. The energy of the states --- the Andreev levels ---
exist in pairs, (one above and one under the
Fermi level), and lie within the
energy gap of the superconductor, with positions which depend on the 
difference  $\phi$
in the phase of the superconductors across the junction, see 
Fig.~\ref{fig:junction}. The applied
bias affects this phase difference through the Josephson relation, 
$\dot{\phi}=2eV/\hbar$. With a constant applied bias $V$ much smaller than the
gap energy $\Delta$, $\phi$ will increase 
linearly in time, and the Andreev levels 
will move adiabatically within the gap.
This motion is a periodic oscillation in $\phi$, indicating that no energy is
transfered to the SQPC and a pure ac current will flow through the contact. 
This is actually the ac Josephson effect.

We wish to study this system in a non-equilibrium situation; one way to  
accomplish this is by introducing microwave radiation with a frequency $\omega\approx \Delta/\hbar$,
which will couple the Andreev levels to each other. The radiation will represent
a non-adiabatic perturbation of the SQPC system. However, if the amplitude 
of the electromagnetic field is sufficiently small, the field will not affect
the adiabatic dynamics of the system much unless the condition for resonant
optical interlevel transitions is fulfilled. Such resonances will only occur 
at certain moments determined by the time evolution of 
the Andreev level spacing. The resonances will provide a mechanism 
for energy transfer to 
the system to be nonzero when averaged over time and hence for 
a finite dc current through the junction. The rate of energy transfer 
is in an essential way
determined by the interference between different scattering events 
\cite{prl,prla,prlf}, and therefore oscillatory 
features appear in the (dc) current-voltage characteristics of an irradiated SQPC. 
Dephasing and relaxation will affect the interference pattern and  
may even conspire to produce a dc current flowing in the reverse direction with
respect to the applied voltage bias. The mechanism behind 
this negative resistance 
is very similar to the one responsible for the ``somersault effect'' discussed 
by Gorelik et al. \cite{somer}.

\section{Theory}
The system we are considering is a one-mode superconducting quantum point contact (SQPC) characterized by an arbitrary energy independent transmission coefficient $D$. A multi-mode junction where a single mode is dominant~\cite{many channels} can also be considered. The junction is weakly biased, $eV\ll\Delta$, and a microwave field is applied to the normal region. The length, $L$, of the normal region is taken to be much shorter than the superconductor coherence length $\xi_0$.
Further, we choose to neglect the proximity effect, which is acceptable in a 
point contact geometry, $W<<\xi_0$, where the current density is low 
except at the very center of the junction, so that the order parameter can be 
assumed to be constant on either side of the contact, which is placed at $x=0$.
As an Ansatz we will work with quasiclassical envelopes $u_\pm(x,t)$, left and right movers, of the two-component wave function $\Psi(x,t)=u_+(x,t)e^{ik_Fx}+u_-(x,t)e^{-ik_Fx}$, as solutions to the time-dependent Bogoliubov-de Gennes equation (BdG)~\cite{book}, 
\begin{equation}
i\hbar \partial{\bf u}/\partial t =[{\bf H_0} + {\bf V_g}(x,t)] {\bf u} \,,
\label{BdG}
\end{equation}
where ${\bf u}=[u_+, u_-]$ is a four-component vector and the gate potential ${\bf V_g}(x,t)=V_\omega(x)\sz\cos \omega t$ induced by the microwave field oscillates rapidly in time. In this text we will use $({\bf u}_1,{\bf u}_2)$ to denote the scalar product between the four-component vectors. The Hamiltonian, ${\bf H}_0$ for the electrons in the electrodes of the point contact,
%
\begin{eqnarray}
\label{H0}
{\bf H}_0&=&-i\hbar v_F\sigma_z\tau_z\partial/\partial x + \\
&&\Delta[\cos(\phi(t)/2)\sigma_x + \sin(\phi(t)/2)\mbox {sgn} x\sigma_y],
\nonumber
\end{eqnarray}
where $\sigma_i$ and $\tau_i$ denote Pauli matrices in
electron-hole space and in $_\pm$ space respectively. 

The function ${\bf u}$, which is smooth on the scale of the Fermi wavelength, 
has a discontinuity at the contact which is determined by the transfer
matrix of the SQPC in the normal state and is described by the following 
boundary condition~\cite{Sh}:
\begin{equation}
{\bf u}(+0)=(1/\sqrt D)(1-\sqrt R\tau_y){\bf u}(-0), \;\; R=1-D,
\label{BC}
\end{equation}
where $D$ is the junction transparency, $0<D<1$.
This form, derived in the appendix, is adequate for our model and is accurate to order $L/\xi_0$.

It is well known that 
there is a discrete set of energy levels within the energy gap of the superconductors in a SQPC. Because of spatial quantization in the transverse direction, the energy spectrum of the Andreev bound states consists of a discrete set of pairs of levels labeled by the quantum number n~\cite{spectra}, 
\begin{equation} E_{n,\pm}=\pm \Delta \sqrt{1-D_n (k_F) \sin^2(\phi/2)}.\label{spectra}
\end{equation}
Energy is measured from the Fermi level and the transmission coefficient $D_n$ is related to the propagation of normal Fermi level electrons
through the microconstriction. The energy of these levels is governed by $\phi=\phi_2-\phi_1$ which is the superconductor phase difference over the junction. 

We are considering a single-mode junction, so $n=1$ and we have two Andreev states, one above and one under the Fermi level, which we label as ${\bf u}^+_\phi(x,t)$ and ${\bf u}^-_{\phi}(x,t)$. 
By assuming that the voltage drop occurs only at the point contact 
we can use the Josephson relation for the phase difference, $\dot{\phi}=2 e V /\hbar$. This assumption together with the restriction that $eV\ll\Delta$, 
allows us to treat ${\bf u}^+_\phi(x,t)$ and ${\bf u}^-_{\phi}(x,t)$ as 
states moving adiabatically in time, through $\phi(t)=2eVt/\hbar$.
This coupling between the phase difference, $\phi$ and time, $t$, allows
us to label the Andreev states as follows, ${\bf u}^+_{\phi(t)}(x)$ and 
${\bf u}^-_{\phi(t)}(x)$.
In this picture we have a pair of energy levels which will oscillate periodically in time with a period of $T_p=\pi\hbar/eV$, see Eq.~\ref{spectra}. These states carry a current when populated and for a junction that is short ($L\ll \xi_0$), all supercurrent through the constriction is carried by these states~\cite{bagwell}.
A study of the boundary conditions for the population is clearly motivated.

\subsection{Boundary condition at $\phi=2\pi n$}
\begin{figure}
\centerline{\psfig{figure=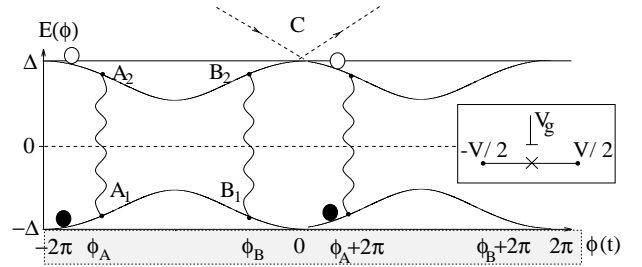,width=8 cm}}
\vspace{3mm}
 \caption{ \label{fig:levels}
Time evolution of Andreev levels (full lines) in the energy gap of a 
gated voltage-biased, single-mode  SQPC (see inset). 
A weak microwave field induces 
resonant transitions (wavy lines) between the levels at points $A$ and $B$ 
and the level above the Fermi energy becomes partly occupied  to an extent 
determined by interference between the two transition amplitudes. 
Non-adiabatic interactions release the energy of quasiparticles in the 
(partly) occupied Andreev level into the continuum 
at point $C$, where the Andreev states and the continuum merge into each other (represented by dashed arrows, see text) and the initial conditions for the 
Andreev level populations are reset (filled and empty circles). 
}
\end{figure}

The interesting region is at $\phi=2\pi n$ (see Fig.~\ref{fig:levels} point C), where the levels approach the continuum and the evolution of states cannot be considered as 
adiabatic, even if we assume a small applied voltage or a weak electromagnetic field. 
Let us study the system in the vicinity of 
$\phi=0$ ($n=0$) and also $t=0$. 

Using the Heisenberg uncertainty relation, $\delta t\delta E\approx\hbar$ and defining $\delta E=\Delta-E(t)=\dot E \delta t$, then assuming $D \approx 1$, we approximate the duration of the non-adiabatic region as,
\be
\delta t=\sqrt{\hbar/\dot E}\approx \hbar (\Delta e^2V^2)^{-1/3}.
\label{t-BC}
\ee
If we want $\delta t\ll T_p$, we find the condition, $eV/\Delta\ll \pi^3$ which means that we can safely treat the non-adiabatic region as short.

The quantity that we are interested in is the transition probability between the state ${\bf u}^+$ at the times $t_1\ll-\delta t$ and $t_2\gg\delta t$. 
Quantum mechanically this can be expressed as the matrix element $\langle({\bf u}^+_{\phi(t_2)},U(t_2,t_1){\bf u}^+_{\phi(t_1)})\rangle$, where $U(t_2,t_1)$ is the exact propagator with our Hamiltonian, Eq.~\ref{H0}. 

%
%

Let us introduce the unitary symmetry operator, ${\bf \Lambda}=\hat P\sx\szz$, under which our Hamiltonian is invariant ($\hat P$ is the parity operator in $x$-space). This property is therefore valid on both sides of $x=0$. An important point now is to show that the boundary condition at $x=0$ does not destroy this symmetry. Let us insert ${\bf \Lambda u}$ into the boundary condition,
\be 
{\bf\Lambda u}(+0)=(1/\sqrt D)(1-\sqrt R\tau_y){\bf\Lambda u}(-0)\\
\Leftrightarrow\hat P\sx\szz{\bf u}(+0)=(1/\sqrt D)(1-\sqrt R\tau_y)\hat P\sx\szz{\bf u}(-0)
\ee
apply $\sx \szz$ from the left and $\hat P$,
\be
{\bf u}(-0)=(1/\sqrt D)(1+\sqrt R\tau_y){\bf u}(+0)
\ee
apply from the left, $(1/\sqrt D)(1-\sqrt R\tau_y)$
\be
(1/\sqrt D)(1-\sqrt R\tau_y){\bf u}(-0)={\bf u}(+0)
\ee
We have recovered the boundary condition (\ref{BC}) which proves that it is invariant under ${\Lambda}$.

Now, we can conclude that both the Hamiltonian, $H_0$, and the boundary condition at $x=0$ are invariant under ${\bf \Lambda}$. This implies that at any time any non-degenerate eigenstate of the Hamiltonian is an eigenstate of ${\bf \Lambda}$ with the eigenvalue $+1$ or $-1$ and that this property persists during the time evolution of the state. Specifically, if we take a state on each side of $\phi=0$,
\be
{\bf \Lambda}{\bf u}^+_\phi&=&\lambda_1 {\bf u}^+_\phi\label{sym1}\\
{\bf \Lambda}{\bf u}^+_{-\phi}&=&\lambda_2 {\bf u}^+_{-\phi}\label{sym2}.
\ee
Next we insert ${\bf u}^+_{-\phi}={\bf \Upsilon} {\bf u}^+_{\phi}$, where the operator ${\bf \Upsilon}=\sx\syy$ changes the sign of the phase $\phi$ into Eq.~\ref{sym2} and apply $\sx\syy$ from the left and arrive at
\be
{\bf \Lambda}{\bf u}^+_\phi=-\lambda_2 {\bf u}^+_\phi,
\ee
which shows that $\lambda_1=-\lambda_2$. This means that the two states, ${\bf u}^+_\phi$ and ${\bf u}^+_{-\phi}$ are orthogonal. 
This is consistent with the results of Shumeiko et.al.~\cite{Sh},
 who have shown that the Andreev state wave functions are 
$4\pi$-periodic whereas the energy levels and the current are
$2\pi$-periodic. 
Since the state evolving from the adiabatic state ${\bf u}^+_{\phi(t_1)}$ is
orthogonal to the adiabatic state ${\bf u}^+_{\phi(t_2)}$,
the probability
for an adiabatic Andreev state to be ``scattered" into a
localized state after passing the non-adiabatic region is identically zero.
In reality,
the Andreev state as it approaches the continuum band edge 
decays into the states of the continuum.  Such a decay corresponds to a
delocalization in real space and is the mechanism for
transferring energy to the reservoir \cite{LV}.

The orthogonality property shown above guarantees that the coherent evolution of our system persists during 
only one Josephson oscillation and that the equilibrium population of the Andreev levels is
reset at each point $\phi=2\pi n$~\cite{Ave}. This imposes the boundary
condition 
\be
b^+(2\pi n+0)=0,\;b^-(2\pi n+0)=1
\label{pop_bc}
\ee
at the beginning of each period.

\subsection{Effect of radiation on the population}
Now we know that we have a system which is periodic, and it is possible to study the effect of the applied high frequency radiation on the population of the Andreev states and the dc-current. We will work with applied high frequency radiation that will induce transitions between the two Andreev levels, described by the eigenfunctions to the Hamiltonian, ${\bf u}^+$ and ${\bf u}^-$. 
The high frequency field  will radically effect the system  when the condition for resonance is fulfilled. This means that even at small amplitudes, $V_\omega \ll \Delta$, when the energy of the field matches the interlevel distance between the two Andreev states, $E_+(t)-E_-(t)=\hbar\omega$ or $2E_+(t)=\hbar\omega$, there will be a large change in their population. 
In this situation we cannot use perturbation theory and instead we will use 
the resonance approximation, where ${\bf u}$ is taken to be a mixture of the 
two states ${\bf u}^+\,\mbox{and}\, {\bf u}^-$,
\begin{eqnarray}
 {\bf u}(x,t)= b^+ {\bf u}^+_{\phi(t)} (x) e^{-i (\omega/2) t}+b^- {\bf u}^-_{\phi(t)} (x) e^{i (\omega/2) t}. \label{res_approx}
\end{eqnarray}
Inserting this Ansatz into the BdG equation~(\ref{BdG}) and
averaging over fast oscillations we arrive at the following coupled differential equation for the population coefficients, $\vec{b}=[b^+, b^-]$;
\begin{equation}
i \dot{\vec{b}}(t)=
\left[ \begin{array}{cc} \delta \omega_{+,-}(t) & V_{+,-}/\hbar
 \\ V_{+,-}^\dagger/\hbar & -\delta \omega_{+,-}(t) \end{array} \right] \vec{b}(t),
\label{eq.coupled}
\end{equation}
where $\delta \omega_{+,-}(t)=[2E_+(t)-\hbar\omega]/2\hbar$, is a measure of the deviation from resonance. $V_{+-}=\int_{-\infty}^\infty dxV_\omega(x)/2(\pcc^+,\sz\p^-)$ is the matrix element between the Andreev states.~\cite{somer}

The matrix above is very close to model Hamiltonians used to study Landau-Zener dynamics of two level systems~\cite{Zener1,Fishman}.
Before we start to study the dynamics of this equation, we need to find an expression for the current through the SQPC. 

\section{Current}
The current carried by populated Andreev levels in equilibrium can be calculated from $I=2 e/\hbar dE_n(\phi)/d\phi$. 
This equation is also valid for adiabatically moving Andreev levels,
however, it needs modification when interaction with a
high frequency field is considered.

We start with the equation for continuity, 
where $\rho_e=e (\pccc,\sz \p)$ is the charge density in a normal conductor ($\Delta = 0$):
\begin{equation}
\partial_t \rho_e+\partial_x I = 0.
\label{tcurr}
\end{equation}
This provides us with the equation for the current, $I(x,t)=-e v_F(\pccc,\szz \p)$.
In superconductors, the charge conservation equation takes form:
\be
&&\partial_t \rho_e+\partial_x I = {\cal I}\\
&&{\cal I} = \frac{2e\Delta}{\hbar} \left[\sin{(\phi/2)}{\rm sgn}(x)
 (\pccc,\sx\p)-\right.\label{Eq:curr11}\\ \nonumber&&\left.\cos{(\phi/2)}(\pccc,\sy\p)\right].
\ee
By multiplying this equation with ${\rm sgn}(x)$ and integrating over the 
whole $x$-axis
we obtain the current at the junction ($x = 0$) as
\begin{equation}
I(0,t) =  \int_{-\infty}^{\infty}dx \,{\rm sgn}(x)\left({\cal I}-\partial_t \rho\right).
\end{equation}
The current $I_{dc}$ averaged over one period of Josephson oscillations reads
\begin{equation}
I_{dc}= \int_{-\infty}^{\infty}dx \,{\rm sgn}(x)\frac{1}{T_p}\left[ \int_0^{T_p} {\cal I}dt - \rho |^{T_p}_0\right].
\label{cur}
\end{equation}
The first term in this equation characterizes the charge exchange between 
Andreev states and condensate. The second term describes the 
ac-component of the current which is connected with a periodic 
charge redistribution in the vicinity of contact. This term is 
equal to zero which is shown in Appendix B.
After a transformation of the quantity $\cal{I}$ (see Appendix B), Eq.~\ref{cur}
can be presented in the equivalent form,
\be
I_{dc}=\frac{2e}{\hbar T_p}\dot{\phi}^{-1}\int_0^{T_p}dt \int_{-\infty}^{\infty}dx \left[
i\hbar \frac{\partial(\pccc,\dot{\p})}{\partial t}-\dot{V}_g (\pccc,\sz\p)\right].
\ee
To find an expression which is valid for our situation with the applied field and with only two energy levels we can insert the resonance approximation, 
Eq.~\ref{res_approx},
\be
I_{dc}=\frac{2e}{\hbar\pi}(\Delta-\frac{\hbar\omega}{2})|\bup(t=T_p)|^2.
\label{eq:curr_b}
\ee
A detailed derivation can be found in Appendix.

To check the validity of the current expression, let us study energy conservation. We have two sources of energy, the applied field and the applied bias. If we consider a single Josephson period, the energy absorbed by the system from the voltage source is $I\cdot V\cdot T_p$ and from the applied field it is $\hbar \omega |\bup(T_p)|^2$. Energy conservation for the period can be stated as, $E_{out}=E_{absorbed}$, with the energy leaving the system stated as $2 \Delta |\bup(T_p)|^2$
\be
2\Delta |\bup(T_p)|^2=I\cdot V\cdot T_p + \hbar \omega |\bup(T_p)|^2,
\ee
the current can easily be found as,
\be
I= \frac{1}{V T_p}(2\Delta |\bup(T_p)|^2-\hbar \omega |\bup(T_p)|^2)
\ee
or, inserting the value of $T_p$,
\be
I=\frac{2e}{\hbar \pi}(\Delta -\frac{\hbar \omega}{2}) |\bup(T_p)|^2.
\ee
This is exactly the same result as the more tedious calculation found in the appendix. So, the current can be understood as voltage bias mediated, photon assisted tunneling.

To sum up, we have shown that the dc-current through a SQPC is linearly 
dependent on the population of the upper Andreev state at the end of the
Josephson oscillation period. This follows from the periodicity of the
population of the Andreev states, Eq.~\ref{pop_bc}. In equilibrium 
without an external field present to shift the population of the Andreev
states, which, at $T=0$, is only in the lower state, there will be no 
current. The introduction of an external field, which is considered here, 
affects the population of the Andreev states in a non-trivial manner and
therefore also the current. The dynamics of the Andreev state populations 
under radiation, governed by Eq.~\ref{eq.coupled}, is therefore the 
next subject.

\section{Dynamics of the Andreev states under radiation}
The current through the SQPC depends linearly on the population of the upper Andreev level at $\phi=2 \pi n^-$, the end of the Josephson period.
Depending on the frequency of the applied field there will be either one, two 
overlapping, or two well separated resonant events during one period. The 
resonances will be Landau-Zener like since the energy levels move in and out
of resonance periodically in time. For a systematic study of the dynamics we 
will
first look at (1) the case of a single resonance and overlapping resonances, 
then we will consider (2) the case of well separated resonances. 

Initially we can discuss two different limits to qualitatively understand 
the dynamics. The first is when the applied bias is large. In this limit 
the energy levels oscillate at a high rate
and there is no time for a resonance to shift the population, the 
population will be untouched. The opposite limit is when the applied 
bias is very small, then the energy levels will oscillate very 
slowly. In this limit the upper level 
will be fully populated and then fully depopulated during one period, once
again the population will be untouched. When the population is untouched
there will be no dc current, and  we can conclude that in these limits
the dc current will be small, and that the interesting region will
be for intermediate bias. The same discussion can be applied for the limits
of a weak and a strong applied external field.



\subsection{Single and overlapping resonances}

\begin{figure}
\centerline{\psfig{figure=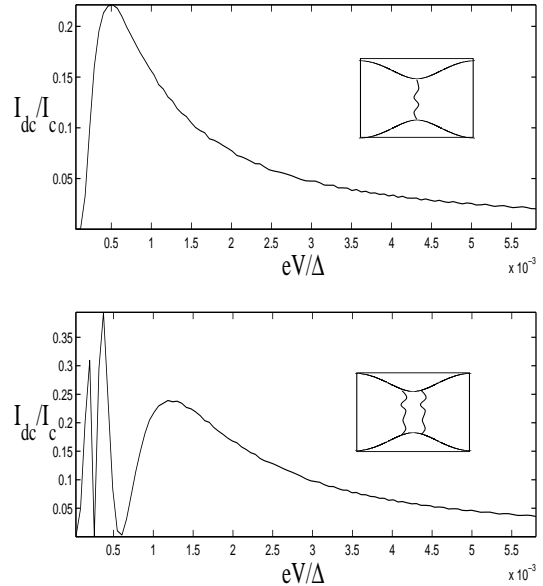,width=7cm,height=8cm}}
 \caption{ \label{fig:currsing}
Numerically calculated, radiation induced current $I_c=e \Delta D/\hbar$. Top, a single resonance and bottom, small deviation from single resonance ($\omega/(\omega_{single}=1.0155$) and the appearance of interference. The time (position) of resonance is shown in the inset figure. In this case the differential equations describing the system have been solved numerically since the time dependence of the energy levels is parabolic and the two resonances can effect each other.  In both plots $\Delta=2K$ and $V_{+,-}/\Delta\approx0.0036$.}
\end{figure}

Consider the case when the applied field matches the interlevel distance in 
the middle of the period, at $\phi=\pi$, see inset of Fig.~\ref{fig:currsing}a.
 At this frequency there is only one resonant event. 
We wish to sweep over a large range of applied bias in order to calculate the current voltage characteristics. 
The term that governs the dynamics is the distance from resonance,
\be 
\delta \omega_{+,-}(t)=\frac{2}{\hbar}(2\Delta \sqrt{1-D_n (k_F) \sin^2(\phi/2)}-\hbar\omega),
\label{eq.dw}
\ee
which has a parabolic time dependence for $\phi\approx\pi$.
This means that we need to solve the system numerically by time iteration of the differential equations~(\ref{eq.coupled}). 

Numerically it is also possible to study the behavior of the current when we first have a single resonance and then in small steps increase the frequency of the applied field producing two resonances, Fig.~\ref{fig:currsing}. The onset of the interference between the two possible paths to the upper levels can clearly be seen. The origin of this interference is that the phase collected by a quasiparticle depends on which level it populates between the two resonant events. This can be compared to the classic double slit experiment in optics.

\begin{figure}
\centerline{\psfig{figure=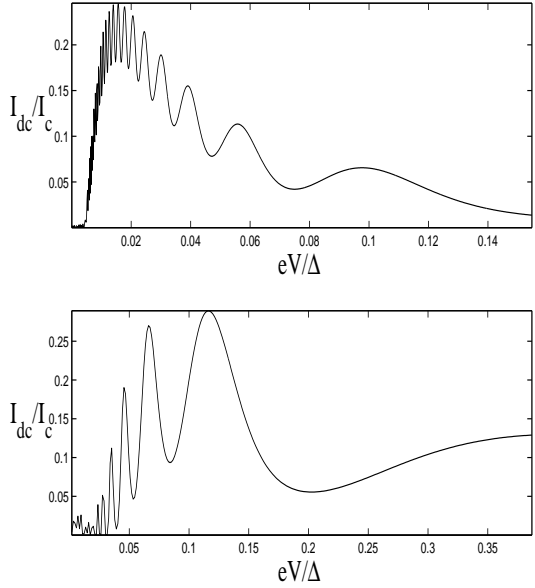,width=7cm,height=8cm}}
 \caption{ \label{fig:single_strongV}
Plot of the single resonance current for an increasing strength of the applied field. In the upper plot, $V_{+.-}/\Delta\approx0.036$ and the lower $V_{+.-}/\Delta\approx0.12$. In both plots $\Delta=3K$.}
\end{figure}

When the strength of the applied field is increased, new oscillations are added to the current, and the envelope of the current is shifted to higher
bias, see the right part of Fig.~\ref{fig:diff_v}. 
We interpret these ``new'' oscillations as a precursor to Rabi oscillations. At
$\phi\approx \pi$, the time dependence is close to zero, $dE/dt\approx0$, and
an analogy can be drawn to a ``long channel'' where Rabi oscillations appear 
between two energy levels with a period depending on the strength of the 
applied field~\cite{tageman}. The population is pumped up and down a number of times during the resonance period resulting in the extra oscillations found in Fig.~\ref{fig:single_strongV}.

\subsection{Separate resonances}

For the case when there are two separate resonances a linear approximation can be applied to the time evolution of the energy levels included in Eq.~\ref{eq.dw}. This is the Landau-Zener model~\cite{LZ} and it is valid  with the restriction that the applied field has to be weak.
Then an asymptotic solution to the differential equations is available,
\be
 d^2=1-e^{-\gamma} \;\;\mbox{and}\;\;
 r^2=1-d^2
\ee
\be \gamma=\pi|V_{+ -}|^2/|dE/dt|,
\ee
where $d^2$ is the probability of changing state after one resonance, and
$r^2$ is the probability of being ``reflected'' back to the same level.
The first step is to construct a unitary scattering matrix for each resonance, $S_A$ and $S_B$, using symmetry arguments which are valid for our system,~Eq.(\ref{eq.coupled}), like $S^\dagger=\sz\, S\, \sz$ and $S_{A,B}=\sx \,S_{B,A}\, \sx$ we have
\be
S_A=\left[ \begin{array}{cc} r&-d e^{-i \Theta} \\ d e^{i \Theta}&  r \end{array} \right],
S_B= \left[ \begin{array}{cc}r &  d e^{i \Theta} \\ 
                 -d e^{-i \Theta}&  r 
 \end{array} \right],
\label{sc_mat}
\ee
where $\Theta$ is the phase that is transferred by a transition. To take care of the phase gathered between the transitions ($\Phi$), we also introduce a propagation matrix
$U=\exp(-i\sz\Phi)$
with
\be
 \Phi&=&\frac{1}{2eV}\int_{\phi_A}^{\phi_B}\left( E_+(\phi)- \frac{\hbar\omega}{2}\right)d\phi.
\end{eqnarray}
The phase $\Phi$ is found through the coupled differential Eqns.~(\ref{eq.coupled}) with $V_{+ -}=0$, the solution far away from resonance,
$\phi_A$ and $\phi_B$ being the two events of resonance, see Fig.~\ref{fig:levels}.

By applying the scattering matrices, Eq.~(\ref{sc_mat}), and the propagation matrix $U$ to the initial state $\vec{b}_0=(0, 1)$ in this fashion $\vec{b}(T_p)=S_B\,U\,S_A\vec{b}_0$, we arrive at
\be
|\bup(t=T_p)|^2=4 r^2 d^2 \sin^2{(\Theta-\Phi)}.
\label{t2theory}
\ee
\begin{figure}
\centerline{\psfig{figure=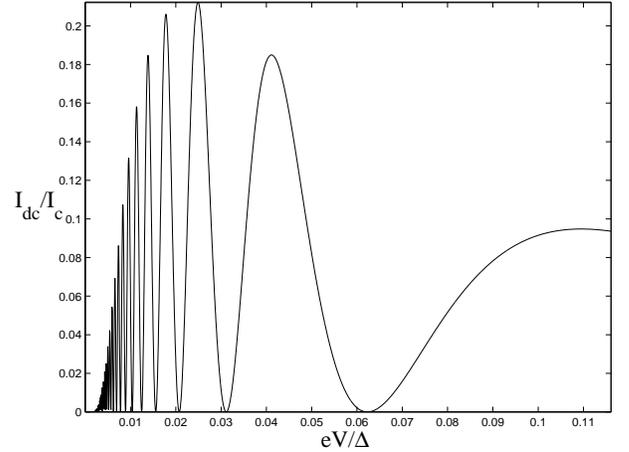,width=8cm}}
 \caption{ \label{fig:currana}
Calculated dc-current, ($I_c=e \Delta D/\hbar$) when the frequency of the applied radiation is such that there are two separate, non-overlapping resonances between the Andreev levels, see Fig. 2. 
In this case the resonances are placed such that the time variation of the energy levels can be approximated as linear and that the two resonances can be treated as separate. $V_{+,-}/\Delta\approx0.048$ and $\Delta=3K$.
}
\end{figure}
The current appears in the subgap region with the following form,  and using Eq.~\ref{eq:curr_b} we acquire,
\be
I_{dc}=\frac{8 e}{\hbar\pi}r^2d^2(\Delta-\frac{\hbar \omega}{2})\sin^2{(\Theta-\Phi).}\label{curr}
\ee
The sine term in Eq.~(\ref{curr}) represents the interference of the two probability amplitudes representing the different paths to the upper level. The phase collected by a particle will be different depending on which level it populates between the two resonances. When the distance between the two resonances is increased the number of oscillations will also increase, see Fig.~\ref{fig:currana} and compare with Fig.~\ref{fig:currsing}b. 
This onset of this interference effect can be studied by comparing Figures~\ref{fig:currsing} and~\ref{fig:currana}. The origin is simple and presents a new method of investigating the nature of Andreev states. The details of this interference is studied and presented in Ref.~\onlinecite{prl}.

\begin{figure}
\centerline{\psfig{figure=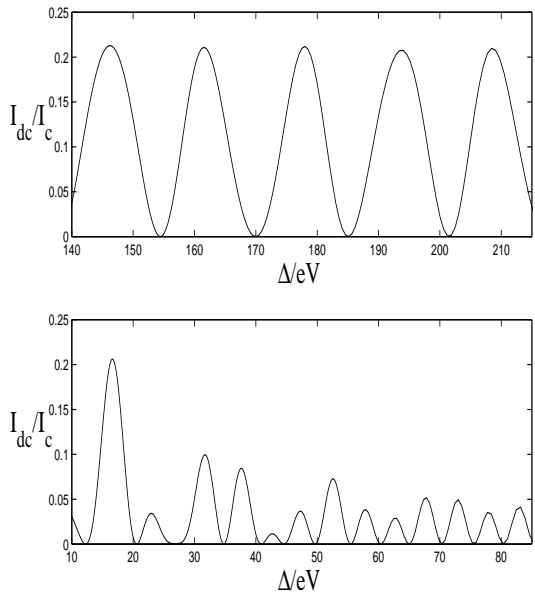,width=7cm,height=8cm}}
 \caption{ \label{fig:diff_v}
Plot of the current as a function of inverse voltage, for a well separated resonance, $\hbar\omega/2\Delta\approx0.76$, with an increasing strength of the applied field. In the upper plot, $V_{+.-}/\Delta\approx0.024$ and the lower $V_{+.-}/\Delta\approx0.12$. In both plots $\Delta=3K$.}
\end{figure}

When the strength of the applied field is increased such that 
$V_{+-}/\Delta\propto 0.1$, and $|\bup|^2$ is calculated 
numerically through Eq.~\ref{eq.coupled},
we find additional oscillations, which the analytical approach failed to
produce. They modulate the $1/V$ periodic oscillations, which have 
their origin in the systems characteristic period $T_p=\hbar\pi/eV$. 
See the left part of Fig.~\ref{fig:diff_v} where the inverse current is 
plotted for three different values of $V_{+-}$. 


When an effort to study the mechanism presented here experimentally is undertaken, the most probable method should be to slowly increase the strength of the applied field and study the I-V curve.
This can be done for several different regimes, (1) for a low or high applied bias and (2) for different frequencies of the applied field, ranging from a single resonance to two well separated ones. 

\begin{figure}[h]
    \centerline{\psfig{figure=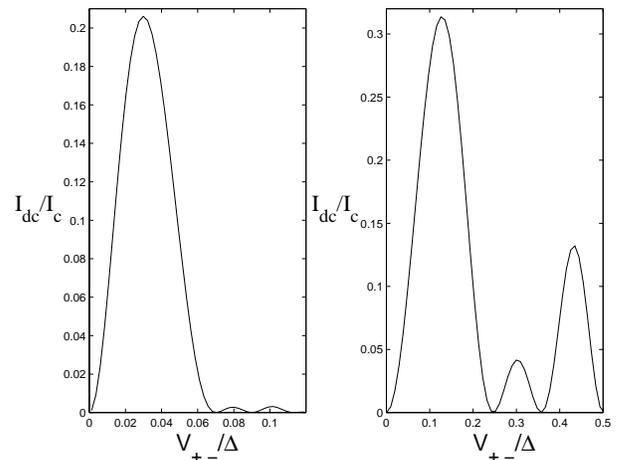,width=8cm}}
    \caption{The current as a function of the applied field strength for the 
case of a single resonance. In the left figure $eV/\Delta\approx 0.012$ 
and in the right figure, $eV/\Delta\approx 0.12$ and in both $\Delta=3K$. 
\label{fig:vsweep}}
\end{figure}

In Fig.~\ref{fig:vsweep} two plots are presented for different values of the applied bias when we have only one single resonance. For the case when $eV/\Delta=0.012$ there appears weak oscillations for higher $V_\pm$. These can probably be understood as Rabi-like oscillations, we can approximate the interlevel distance as constant during the resonance and the population can be pumped up and down several times. 

\begin{figure}[h]
    \centerline{\psfig{figure=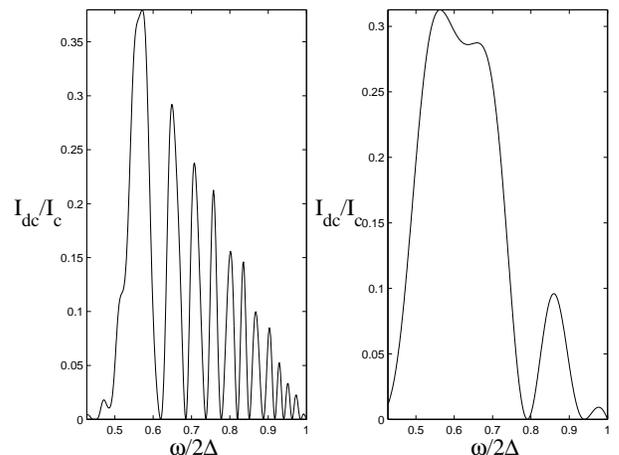,width=8cm}}
    \caption{The current as a function of the frequency of the applied field.
    In the left figure $eV/\Delta\approx 0.019$ and $V_{+,-}/\Delta\approx 
0.036$, while in the right figure, $eV/\Delta\approx 0.12$ and 
$V_{+,-}/\Delta\approx 0.12$ In both plots $\Delta=3K$. \label{fig:wsweep}}
\end{figure}

Another case that can be studied is the current as a function of the applied field's frequency, $\hbar\omega$. The familiar oscillation will appear even in this case, see Fig.~\ref{fig:wsweep}. By slowly increasing the frequency of the applied field and thus shifting the position of the resonances, the  sine term in Eq.~\ref{curr} is modulated.

\subsection{Relaxation and dephasing}

Up to this point we have studied an ideal system, assuming that  Landau-Zener like transitions dominate. We have neglected the effect of voltage fluctuations and interactions with external degrees of freedom. 
If we assume that the relaxation and dephasing times, $\tau, \tau_\phi$ are short compared to the duration of the non-adiabatic resonances, $\delta t$, Eq.~\ref{t-BC}, we can phenomenologically study the effect of such interactions on our system. The technique~\cite{Gefen} we will use models dissipation by adding a term to the time evolution equations of the density matrix for our two-level system. The equations have the following form,
\be
&&\dot \rho_{nn}(t)=-\frac{i}{\hbar}[H_0(t),\rho(t)]_{nn}-\frac{\rho_{nn}-\rho_{nn}^{eq}}{\tau},\label{eq.d1} \\
&&\dot \rho_{nn'}(t)=-\frac{i}{\hbar}[H_0(t),\rho(t)]_{nn'}-\frac{\rho_{nn'}}{\tau_{\phi}}, \; \mbox{n$\neq$n'},\label{eq.d2}
\ee
with $\tau$ as the characteristic time for relaxation of the system to the equilibrium population, $\rho^{eq}$, and  $\tau_\phi$ as the characteristic time for dephasing.

The exact form of the density matrix for the system considered here will be a $2\times2$ matrix for the discrete two level system of the Andreev states (To avoid confusion we will use $\sxxx$, $\syyy$, $\szzz$ to denote the Pauli matrices in this discrete space). The Hamiltonian for the density matrix is found in Eq.~\ref{eq.coupled}., and the diagonal elements $\rho_{11}$ and $\rho_{22}$ will represent the population if the upper and lower Andreev levels. The resonant events which are unaffected by the dissipation can still be modeled with the scattering matrices introduced earlier, Eq.~\ref{sc_mat},
\be
\rho'=\hat{S}_{A,B}\; \rho\; \hat{S}_{A,B}^\dagger.
\ee

To demonstrate 
the effect of relaxation and dephasing we will calculate
the dc-current for well separated resonances.
In this case the system is strongly effected by the high frequency field
only in the vicinity of $t=t_{1,2}$. Therefore 
during almost the whole interval of time, $T_p$, one can neglect 
${\bf V_{g}}$ in Eq.(1) and calculate the 
instantaneous current $I(t)$ by using the expression
\be
I(0,t) = \frac{2e}{\hbar} \dot{\phi}^{-1} \frac{d \,{\rm Tr}({\bf H}_0 \rho )}{dt}.
\label{eq.incur}
\ee  
This expression is an extension of Eq.(\ref{cur}) into the case of relaxation.
As a result in adiabatic limit $\hbar \dot{\phi}<<\Delta$,
the dc-current is given by 
\be
I_{dc}=&&\frac{1}{T_p}\frac{2e}{\hbar}\int_0^{T_p} {\rm Tr}(\szzz \rho)\partial E/\partial\phi\; dt=\nonumber\\&&\frac{1}{T_p V}\int_0^{T_p} \frac{\partial E_+(t)}{\partial t}\left[2 \rho_{11}-1\right]dt,
\label{eq.dispcurr}
\ee
where we have used the Josephson relation $\dot{\phi}=2eV/\hbar$ and that $Tr (\rho)=1$. 

By solving the time evolution of the density matrix for the whole Josephson period and imposing the boundary condition $\rho^{eq}=(1-\szzz)/2$ at $t=0$, we arrive at the following,

\be
\begin{raggedleft}
\rho_{11}=
\left\{
\begin{array}{lll}
0& \text{, $0< t< t_1$}& \\
d^2\; e^{-(t-t_1)/\tau}& \text{, $t_1< t < t_2$}&\\
f(r^2,\tau_\phi,\tau)\;e^{-(t-t_2)/\tau} &\text{, $t_2 < t < T_p$}&
\end{array} \right.
\end{raggedleft}
\ee
with
\be
f(r^2,\tau_\phi,\tau)=&&\left[ d^2 (1-e^{-(t_2-t_1)/\tau})+
2r^2 d^2 (e^{-(t_2-t_1)/\tau}-\right.\nonumber\\
&&\left.e^{-(t_2-t_1)/\tau_\phi} \cos{2(\Theta+\Phi)})\right],
\label{eq.f}
\ee
where $d^2=1-r^2$.
Inserting these expressions into Eq.~\ref{eq.dispcurr} we find,
\be
&&I_{dc}(V)=\frac{2}{VT_p}\left[ \frac{d^2\hbar\omega}{2}\left(e^{-(t_2-t_1)/\tau}-1\right) + \frac{d^2}{\tau}\int_{t_1}^{t_2}E_+ (t) dt + \right.\nonumber\\
\nonumber&&\left. f(r^2,\tau_\phi,\tau) \left\{\left(\Delta e^{-(T_p-t_2)/\tau} -\frac{\hbar\omega}{2}\right)+\right.\right.\\&&\left.\left. \frac{1}{\tau} \int_{t_2}^{T_p}E_+(t) e^{-(t-t_2)/\tau} dt \right\}\right].
\ee

\begin{figure}[h]
        \centerline{\psfig{figure=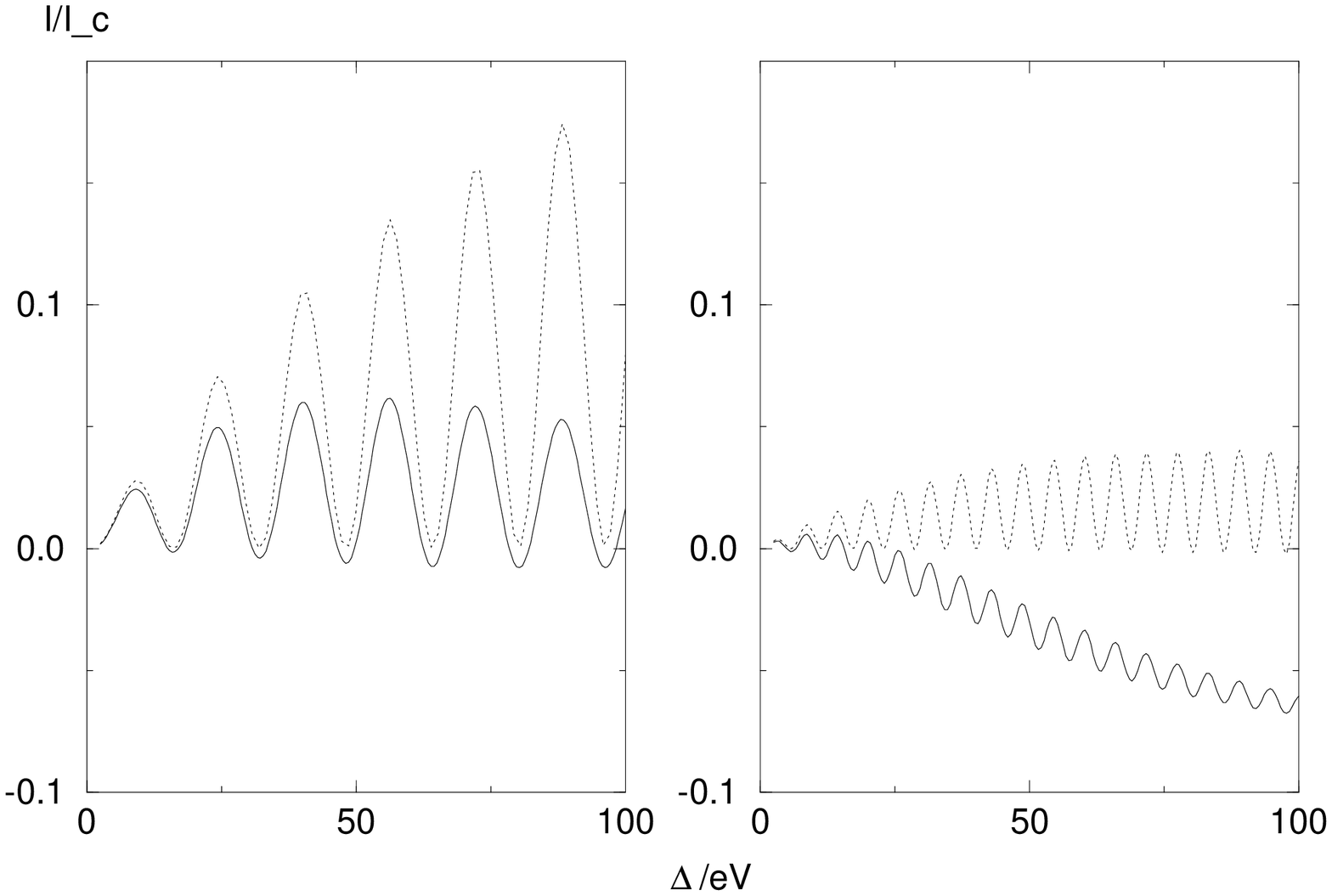,width=8cm}}
    \caption{The dc current calculated with dephasing and dissipation 
phenomenologically added. In both cases $\tau_\phi\approx\tau\approx 
4\cdot 10^{-10}s$. The frequency of the applied field is such that $\hbar\omega
/2\Delta \approx0.75$ in the left plot $\hbar\omega /2\Delta \approx0.95$ in 
the right. Note the fact that the system shows a {\it negative conductance}
in the right plot. The amplitude of the applied field is such that $|V_{+-}|
/\Delta \approx 0.025$ in both plots, and for reference the case when
$\tau,\tau_\phi\rightarrow\infty$ are included (dotted curves).\label{fig:diss}}
\end{figure}

This formula 
falls back into the previous current expression that we have derived, Eq.\ref{curr}
when $\tau_\phi\rightarrow\infty$ and $\tau\rightarrow\infty$. 
From expression (\ref{eq.f}) one can see 
that the oscillations of the current 
exponentially decrease when the dephasing time $\tau_\phi$ becomes smaller
than $T_p$. 
At the same time when the relaxation time $\tau$ is decreased the current 
drops to zero as $\tau$. 
The I-V curves for the case $ T_p < \tau_\phi\approx 2\tau$ ~\cite{Gefen}
are presented in Fig.~\ref{fig:diss}.

A curious effect occurs when the relaxation time is such that the upper level is fully depopulated when $\phi=\pi$, in the middle of the Josephson period. The result is a negative current when the dc current is calculated for a positive bias, our system could under special circumstances show a {\em negative conductance} see Fig.~\ref{fig:diss}. The reason is that the upper level is mainly populated when it's derivative is negative and then the lower level is highly populated for the part of the period when it's derivative is negative. The effect should be most pronounced when the energy of the applied field, $\hbar \omega$, approaches the energy of the gap, $2\Delta$. 


Actually. the physical mechanism which is behind the appearance of this negative resistance is similar one discussed in~\cite{somer}.

\section{Conclusions}
We have shown that irradiation of a voltage-biased
superconducting quantum point contact at frequencies $\omega\sim\Delta$ can 
remove the suppression of subgap dc transport through Andreev levels. 
Quantum interference among resonant scattering events can be used
for microwave spectroscopy of the Andreev levels. 

The same interference effect can also be applied for detecting
weak electromagnetic signals up to the gap frequency. Due to the resonant
character of the phenomenon, the current response is proportional to the
ratio between the amplitude of the applied field and the applied voltage,
$I\sim|V_\pm|^2/\Delta eV$. At the same time,
for common SIS detectors a non-resonant current response 
is proportional to the ratio between the amplitude and the frequency
of the applied radiation \cite{Tucker}), $I\sim |V_{\pm}/\omega|^2$, i.e. 
it depends entirely on the parameters of the external signal
and cannot be improved.

A large part of the text is devoted to a thorough study of the dynamics
of adiabatic Andreev levels in a biased SQPC
subject to microwave radiation. We have presented calculations of the
dc-current where we vary either the frequency or the strength of the microwave field, or the voltage bias.   
These results are open for experimental investigation since 
a shift in the population of the Andreev states is directly
reflected in the current through the SQPC.

Finally, we note that
the classic double-slit interference experiment, where two spatially separated
trajectories combine to form an interference 
pattern, clearly demonstrates the wave-like nature of electron propagation.
For a 0-dimensional
system, with no spatial structure, we have shown that a completely 
analogous interference phenomenon
may occur between two distinct trajectories in the  {\em temporal} 
evolution of a quantum system. 
%
 
\section{Acknowledgment}
Support from the Swedish KVA, SSF, Materials consortia 9 \&
11, NFR and from the National Science Foundation under Grant No. 
PHY94-07194 is gratefully acknowledged. 

\appendix
\section{Boundary condition at $x=0$}

\begin{figure}
\centerline{\psfig{figure=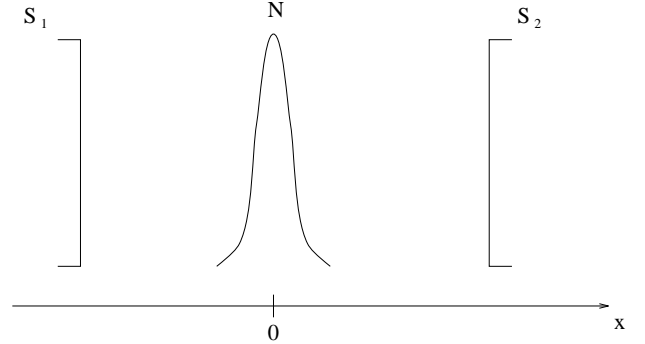,width=8cm}}
\caption{\label{fig:sns}
Schematic for SNS junction with scatterer.}
\end{figure}

Here we are going to calculate a boundary condition for ${\bf u}\pm(x,t)$ at the
normal region with a scatterer. This is done under the assumption that we have one scatterer located at $x=0$ and that $\Delta\ll \mu$.

In the left superconductor we make the Ansatz,
\be
\Psi_1=
\left(\begin{array}{cc}
a_1\\b_1\end{array}\right) e^{i k x} + 
\left(\begin{array}{cc}
c_1\\d_1\end{array}\right) e^{-i k x}
\ee
and analogously for the right superconductor,
\be
\Psi_2=
\left(\begin{array}{cc}
a_2\\b_2\end{array}\right) e^{i k x} + 
\left(\begin{array}{cc}
c_2\\d_2\end{array}\right) e^{-i k x}.
\ee

\begin{figure}
\centerline{\psfig{figure=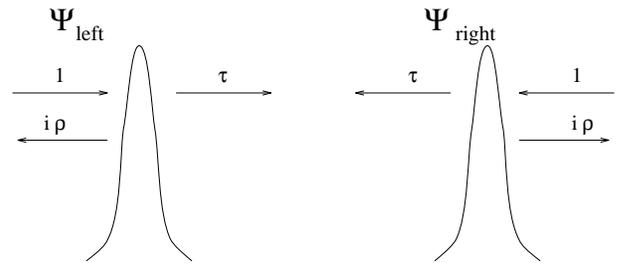,width=8cm}}
\caption{Scattering states, from left and right.}
\label{fig:scatter}
\end{figure}

The Ansatz for the normal region will be two linear combinations of the
possible scattering states, see fig.~\ref{fig:scatter},
\be
\left( \begin{array}{cc}
A \Psi_{left} + B \Psi_{right}\\
C \Psi_{left} + D \Psi_{right}
\end{array} \right).
\ee
Different parts of this Ansatz exists to the left and the right side of the scatterer;
\newline
for $x<0$:
\be
\Psi_N=
\left(\begin{array}{cc}
A\\C\end{array}\right) e^{i k x} + 
\left(\begin{array}{cc}
A i \rho + B \tau\\C i \rho + D\tau\end{array}\right) e^{-i k x}
\ee
and for $x>0$:
\be
\Psi_N=
\left(\begin{array}{cc}
A\tau+B i \rho\\C\tau+D i \rho\end{array}\right) e^{i k x} + 
\left(\begin{array}{cc}
B\\D \end{array}\right) e^{-i k x}
\ee

By matching through this structure, Fig.~\ref{fig:sns} with the Ansatzes as defined above, we arrive at the following condition,
\be
\left( \begin{array}{cc}
a_1 \\ b_1 \\ c_1 \\ d_1 \end{array}\right)=
1/\tau \left( \begin{array}{ccccc}
1 & 0 & -i \rho & 0 \\
0 & 1 & 0 & -i \rho \\
i \rho & 0 & 1 & 0 \\
0 & i \rho & 0 & 1 
\end{array}\right)
\left( \begin{array}{cc}
a_2 \\ b_2 \\ c_2 \\ d_2 \end{array}\right).
\ee
Using the notation with Pauli matrices we can write this boundary condition as,
\be
\Psi(-x)=\frac{1}{\sqrt{D}}(1+\syy\sqrt{R})\Psi(x)
\ee
or,
\be
\Psi(x)=\frac{1}{\sqrt{D}}(1-\syy\sqrt{R})\Psi(-x).
\ee

\section{Current expression for a biased junction}
In this appendix we derive an expression for the time-dependent current through an SNS structure when a weak bias is applied. An applied bias forces the Andreev levels to oscillate introducing a time-dependence, which makes the validity of the commonly used expression, $I=(2e/\hbar)\partial E/\partial \phi$, questionable. 
 
We wish to find an expression for the current that includes time-dependence. The time-dependence enters both through the phase difference $\phi(t)=2eVt/\hbar$ and the applied field $V_g(t)$. 

According to Eq.~\ref{Eq:curr11}, charge conservation in the superconductor reads:
\be
\hbar (\partial_xI+\partial_t \rho_e) =  2e\Delta \left[\sin{(\phi/2)}{\rm sgn}(x) (\pccc,\sx\p)- \right.\label{Eq:curr111} \\
\left. \nonumber\cos{(\phi/2)}(\pccc,\sy\p) \right],
\ee
Using the identity
\be
i\hbar\partial_t(\pccc,\dot\p)=(\pccc,\dot H\p),
\ee
which directly follows from the BdG equation (1), we find

\be
I(0,t)=\frac{2e}{\hbar}\dot{\phi}^{-1}\int_{-\infty}^{\infty}dx \left[
i\hbar \frac{\partial(\pccc,\dot\p)}{\partial t}-\dot{V}_g (\pccc,\sz\p)\right] \nonumber\\
-\frac{1}{2}\partial_t\left(\int_{-\infty}^0\rho dx -\int_0^{\infty}\rho dx \right).
\label{Eq:currQ}
\ee

To calculate the charge density $\rho_e$, one needs to examine the properties of
the Andreev state wave functions. 
This is done by evaluating the expression for the
charge density $\rho_e = e({\bf u},\sigma_z {\bf u})$
with the resonance approximation Eq.~(\ref{res_approx}) inserted, 
the result is a  linear combination of the terms
$({\bf u}^{\pm},\sigma_z {\bf u}^{\pm})$
and $({\bf u}^{\pm},\sigma_z {\bf u}^{\mp})$.
It follows from Eq.(\ref{H0}) that the 4-vectors ${\bf u}^{\mp}$ satisfy
the symmetry equation ${\bf u}^{\mp} = 
\tau_z\sigma_x \left({\bf u}^{\mp}\right)^\ast$
. This immediately gives that $({\bf u}^{\pm},\sigma_z {\bf u}^{\pm}) = -
({\bf u}^{\pm},\sigma_z {\bf u}^{\pm})^\ast = 0$.
Furthermore , at $|E|=\Delta$ ($t = 0,T_p$), 
the Andreev state wave function in a contact with
normal electron scattering ($R \not= 0$) has a form similar to
the wave functions in a perfect constriction ($R= 0$) (Ref[7]).
or the case of a perfect constriction, the Andreev state 
wave functions ${\bf u}^{+}$ and
${\bf u}^{-}$, are  eigenfunctions of $\tau_z$ with different 
eigenvalues ($\pm 1$).
This further gives that $({\bf u}^{\pm}_{\phi=2\pi n}, 
\sigma_z {\bf u}^{\mp}_{\phi=2\pi n}) = 0$

Therefore the appropriate current expression is,
\begin{eqnarray}
I_{dc}&=&\frac{1}{T}\int_0^{T_p}I(0,t)dt=\\
&=&\frac{2e}{\hbar T_p}\dot{\phi}^{-1}\int_0^{T_p}dt \int_{-\infty}^{\infty}dx \left[ i\hbar \frac{\partial(\pccc,\dot{\p})}{\partial t}-\dot{V}_g (\pccc,\sz\p)\right],
\ee
or
\be
I_{dc}=
\frac{2e}{\hbar T_p}\dot{\phi}^{-1}\left[ i \hbar \langle(\pccc,\dot{\p}\rangle|^{T_p}_0-\int_0^{T_p}dt \langle\dot{V}_g (\pccc,\sz\p) \rangle \right],
\ee
where $\langle..\rangle$ symbolizes integration over the whole x-axis.

\section{Current expression under irradiation}
The result in the previous appendix is a general expression for the dc current through a short biased SQPC. Here we will specialize the current expression to the case of an irradiated single mode junction. 

For the first part $\langle(\pccc,\dot{\p})\rangle|^{T_p}_0$  we are only interested in the end points of the cycle, $t=0$ and $t=T_p$. At these two occasions the wave functions are well defined, a mixture of the two available states which are orthogonal.
\be
\p=\bup(t) \ffiup (x) e^{-\frac{i E_{\uparrow}t}{\hbar}}+\bner(t) \ffiner (x) e^{-\frac{i E_{\downarrow}t}{\hbar}}
\ee
We also need the time derivate, $\dot{\p}$, where we can neglect $\dot{\bup}$ and $\dot{\bner}$ since the population is stable far from resonance.
\be
\dot{\p}=-\frac{i E_{\uparrow}}{\hbar}\bup(t) \ffiup (x) e^{-\frac{i E_{\uparrow}t}{\hbar}}-\frac{i E_{\downarrow}}{\hbar}\bner(t) \ffiner (x) e^{-\frac{i E_{\downarrow}t}{\hbar}}
\ee
The following is known; at $t=0$ and $t=T_p$: $E_{\uparrow}=\Delta$ and $E_\downarrow=-\Delta$, and that at $t=0$: $\bner=1$ and $\bup=0$.
So we have,
\be
(\pccc,\dot{\p})=-\frac{i \Delta}{\hbar}(|\bup|^2 |\ffiup|^2-|\bner|^2 |\ffiner|^2)
\ee
and
\be
\langle(\pccc,\dot{\p})\rangle=-\frac{i \Delta}{\hbar}(|\bup|^2 -|\bner|^2 )
\ee
or, using the normalization, $|\bup|^2 +|\bner|^2=1$,
\be
\langle(\pccc,\dot{\p})\rangle=-\frac{i \Delta}{\hbar}(2 |\bup|^2 -1).
\ee
Now we can express the first part of the current as,
\begin{eqnarray}
&&\frac{i e}{\pi} (-\frac{i \Delta}{\hbar}(2 |\bup|^2 -1))|_0^{T_p}= \label{eq.p1c}\\
&&\frac{e\Delta}{\hbar\pi}(2 |\bup(t=T_p)|^2 -1 -(2 |\bup(t=0)|^2-1)=
\nonumber\\ 
&&\frac{2e\Delta}{\hbar\pi} \left(|\bup(t=T_p)|^2-|\bup(t=0)|^2\right)\nonumber
\end{eqnarray}
The second part is a little more complicated, here we have to use the resonance approximation for the wave function and insert the applied potential $V_g(x,t)=V_\omega(x) \cos{(\omega t)}$. Then we find,
\begin{eqnarray}
&&\langle\dot{V}_g(x,t)(\pccc,\sz\p)\rangle=\nonumber\\
&&[\mbox{Omitting oscillating terms and defining } \nonumber\\
&&V_{+-}=\langle\frac{V_\omega}{2}(\pccc^+, \sz \p^-) \rangle]=\label{eq.p2}\\
&&\nonumber\frac{\omega}{i} (\bner V_{+-} (\bup)^\dagger-(\bner)^\dagger V_{+-}^\dagger \bup)=\\ \nonumber
&&\omega ({\rm Re}[V_{+-}] \vec{b}^\dagger\syyy\vec{b}+{\rm Im}[V_{+-}] \vec{b}^\dagger\sxxx\vec{b}),
\end{eqnarray}
where $\vec{b}=[\bup, \bner]$.

Since we need to integrate this expression over one cycle, we need more information. This can be found from the coupled differential equations governing the behavior of $\vec{b}$. If we rewrite these equations using Re[\,] and Im[\,] we get the following form
\be
i \dot{\vec{b}}=-\frac{\delta \omega}{2}\szzz\vec{b}+\frac{{\rm Re}[V_{+-}]}{\hbar}\sxxx\vec{b}-\frac{{\rm Im}[V_{+-}]}{\hbar}\syyy\vec{b}.
\label{eq.pp}
\ee
Next, if we operate from the right with $\vec{b}^\dagger \szzz$ and on the conjugate of the above equation with $\vec{b} \szzz$ we get the following two equations,
\be
i \vec{b}^\dagger\szzz\dot{\vec{b}}=\frac{\delta \omega}{2}\vec{b}^\dagger\vec{b}+\frac{{\rm Re}[V_{+-}]}{\hbar}i\vec{b}^\dagger\syyy\vec{b}+\frac{{\rm Im}[V_{+-}]}{\hbar}i\vec{b}^\dagger\sxxx\vec{b}
\ee
and
\be
-i \vec{b}\szzz\dot{\vec{b}}^\dagger=\frac{\delta \omega}{2}\vec{b}^\dagger\vec{b}-\frac{{\rm Re}[V_{+-}]}{\hbar}i\vec{b}^\dagger\syyy\vec{b}-\frac{{\rm Im}[V_{+-}]}{\hbar}i\vec{b}^\dagger\sxxx\vec{b}.
\ee
Finally, take the difference between the two,
\be
i (\vec{b}^\dagger\szzz\dot{\vec{b}}+\vec{b}\szzz\dot{\vec{b}}^\dagger)=\frac{2i}{\hbar}({\rm Re}[V_{+-}]\vec{b}^\dagger\syyy\vec{b}+{\rm Im}[V_{+-}]\vec{b}^\dagger\sxxx\vec{b}).
\ee
Then we have on the right hand side, part of Eq.~\ref{eq.p2} and combining these two we obtain,
\be
i\partial_t (\vec{b}^\dagger\szzz\vec{b})=\frac{2i}{\hbar\omega}\langle\dot{V}_g(\pccc,\sz\p)\rangle
\ee
or
\be
\langle\dot{V}_g(\pccc,\sz\p)\rangle=\frac{\hbar\omega}{2}\partial_t (\vec{b}^\dagger\szzz\vec{b}).
\ee
the second part of the current expression now follows,
\begin{eqnarray}
&&\frac{e}{\hbar\pi}\int_0^{\hbar\pi/eV}\langle\dot{V}_g(\pccc,\sz \p)\rangle=\frac{e}{\hbar\pi}\int_0^{\hbar\pi/eV}\frac{\hbar\omega}{2}\partial_t (\vec{b}^\dagger\szzz\vec{b})=\nonumber\\
&&\frac{e\omega}{2\pi}(\vec{b}^\dagger\szzz\vec{b})|_0^{T_p}=\frac{e\omega}{2\pi}(|\bup|^2-|\bner|^2)|_0^{T_p}=\frac{e\omega}{2\pi}(2|\bup|^2-1)|_0^{T_p}=\nonumber\\
&&\frac{e\omega}{\pi}\left(|\bup(t=T_p)|^2-|\bup(t=0)|^2\right).\label{eq.p2c}
\end{eqnarray}

Putting the two parts together, \ref{eq.p1c}-\ref{eq.p2c}, the current will be,
\be
I_{dc}(x=0)=\frac{2e}{\hbar\pi}\left(\Delta-\frac{\hbar\omega}{2}\right)
\left[ |\bup(t=T_p)|^2- |\bup(t=0)|^2\right].
\nonumber
\ee


\begin{references}

\bibitem{Altshuler}
B.L.Altshuler and A.G.Aronov, in Modern Problems in Condensed
Matter Science, edited by A.L.Efros and M.Pollak, Amsterdam (1985)

\bibitem{Datta}
S.Datta, Electronic transport in mesoscopic systems, edited by
H.Ahmed, M.Pepper, and A.Broers (cambridge University Press, 1995)

\bibitem{HN}
F.Hekking and Yu.V.Nazarov, Phys.Rev,B {\bf 44}, 11506 (1991)

\bibitem{somer} L. Y. Gorelik, V. S. Shumeiko, R. I. Shekhter, G. Wendin and M. Jonson, Phys. Rev. Lett. {\bf 75}, 1162 (1995).

\bibitem{prla}
L.Y.Gorelik, A.Grincwajg, V.Z.Kleiner, R.I.Shekhter, and M.Jonson,
Phys. Rev. Lett. {\bf 73}, 2260 (1994)

\bibitem{prlf}
L.Y.Gorelik, F.A.Maa\o,  R.I.Shekhter, and M.Jonson,
Phys. Rev. Lett. {\bf 78}, 3169, (1997)

\bibitem{GT}
Y.Gefen and D.J.Thouless, Phys.Rev.Lett. {\bf 59}, 1752, (1987).

\bibitem{prlg}
L.Y.Gorelik,  S.Kulinich, Y.Galperin, R.I.Shekhter, and M.Jonson,
Phys. Rev. Lett. {\bf 78}, 2196 (1997)

\bibitem{prl} L. Y. Gorelik, N. I. Lundin, V. S. Shumeiko, R. I. Shekhter and M. Jonson, Phys. Rev. Lett. {\bf 81}, 2538 (1998).

\bibitem{many channels} E.~Scheer, P.~Joyez, D.~Esteve, C.~Urbina and M.~H.~Devoret, Phys. Rev. Lett. {\bf 78}, 3535, (1997).

\bibitem{kulik} I. O. Kulik and A. N. Omel'yanchuk, Fiz. Nisk. Temp. {\bf 3}, 945 (1977);{\bf 4}, 296 (1978) [Sov. J. Low Temp. Phys. {\bf 3}, 459 (1977); {\bf 4}, 142 (1978)].

\bibitem{book} P. G. de Gennes, {\em Superconductivity of Metals and Alloys} 
(Addison-Wesley, New York, 1966).

\bibitem{Sh} V.S. Shumeiko, G. Wendin, and E.N. Bratus', Phys. Rev. B {\bf 48}, 13129 (1993); V.S. Shumeiko, E.N. Bratus', and G. Wendin, Low Temp. Phys. {\bf 23}, 181 (1997).

\bibitem{spectra} C.~W.~J.~Beenakker, Phys. Rev. Lett. {\bf67}, 3836 (1991).

\bibitem{bagwell} P. F. Bagwell, Phys. Rev. B {\bf 46}, 12 573, (1992).

\bibitem{LV}
This conclusion is in agreement with the analysis of multiple Andreev
reflections in Ref.~\cite{gapcurr}.

\bibitem{Ave} D.~Averin and A.~Bardas, Phys. Rev. B {\bf 53}. R1705, (1996).

\bibitem{Zener1} D.~Bouwmeester, N.~H.~Dekker, F.~E.~v.~Dorsselaer, C.~A.~Shrama, P.~M.~Visser, and J.~P.~Woerdman, Phys. Rev. A {\bf 51}, 646 (1995).

\bibitem{Fishman} S. Fishman, K. Mullen and E. Ben-Jacob, Phys. Rev. A {\bf 42}, 5181 (1990).

\bibitem{tageman}
O. Tageman, L. Y. Gorelik, R. I. Shekhter and M. Jonson,
J. Appl. Phys. {\bf 81}, 285-291, (1997).

\bibitem{LZ} B. W. Shore, {\it The Theory of Coherent Atomic Excitations} (Wiley, New York 1990), Vol. 1.

\bibitem{Gefen}
E.~Shimshoni and Y.~Gefen, Ann. Phys. {\bf 210}, 16 (1991).

\bibitem{Tucker}
J. R. Tucker and M. J. Feldman, Rev. Mod. Phys. {\bf 57}, 1055 (1985).

\bibitem{gapcurr} E.~N.~Brutus', V.~S.~Shumeiko, E.~V.~Bezuglyi and G.~Wendin, Phys. Rev. B {\bf 55}, 12 666, (1997).





\end{references}
\end{document}